# SAR Image Segmentation using Vector Quantization Technique on Entropy Images

Dr. H. B. Kekre
Computer Engineering
MPSTME, NMIMS University,
Vileparle(w)
Mumbai 400–056, India
hbkekre@yahoo.com.

Saylee Gharge
Ph.D. Scholar, MPSTME,
NMIMS University,
Assistant Professor, V.E.S.I.T,
Mumbai-400071, India
sayleegharge73@yahoo.co.in

Tanuja K. Sarode
Ph.D. Scholar, MPSTME,
NMIMS University,
Associate Professor, TSEC,
Mumbai 400-050, India
tanuja_0123@yahoo.com

*Abstract*—**The development and application of various remote sensing platforms result in the production of huge amounts of satellite image data. Therefore, there is an increasing need for effective querying and browsing in these image databases. In order to take advantage and make good use of satellite images data, we must be able to extract meaningful information from the imagery.**

**Hence we proposed a new algorithm for SAR image segmentation. In this paper we propose segmentation using vector quantization technique on entropy image. Initially, we obtain entropy image and in second step we use Kekre's Fast Codebook Generation (KFCG) algorithm for segmentation of the entropy image. Thereafter, a codebook of size 128 was generated for the Entropy image. These code vectors were further clustered in 8 clusters using same KFCG algorithm and converted into 8 images. These 8 images were displayed as a result. This approach does not lead to over segmentation or under segmentation. We compared these results with well known Gray Level Co-occurrence Matrix. The proposed algorithm gives better segmentation with less complexity**

***Keywords-component; SAR image; image Segmentation; Probability; Entropy; Vector Quantization; Codevector;.***

## I. INTRODUCTION

The synthetic aperture radar (SAR) system has been of great use in monitoring the global environment, observing land usage, investigating disaster regions [1,2,3,4] as well as detecting military targets in the early days of the system [5, 6, 7]. One of the important role of SAR system is to collect the information about the ground surface through image reconstruction. Since the ground surface has diverse areas, such as rice fields, wheat fields, grassland, ponds, asphalt roads, desert, and so forth, those areas have to be segmented in the image reconstruction. Thus segmentation problem arises. Because of these, various segmentation methods have been proposed in the framework of image processing. In recent research work, a method based on computing optimal threshold by entropy maximization is derived in [8], the combination of multilevel logistic model and EM algorithm yields an unsupervised segmentation method as proposed in [9], Computer-assisted algorithms to segment SAR sea ice imaginary are discussed in [10], oil spill segmentation is considered by using minimum description length and a polynomial active grid in [11], and a segmentation method for sea ice SAR imaginary based on pulse coupled neural networks is presented in [12]. Furthermore, hybrid segmentation schemes are derived by combining Wishart segmenter and so-called H/A/alpha segmenter in [13], and multiscale Bayesian algorithm based on hierarchical Markov random field is proposed in [14].

High resolution sensor technique has made great progress since 1990's. High resolution image can show the object information such as structure, texture and detail clearly. Texture feature is the direct embodiment of the object structure and space arrangement in the image. Recently, resolution of remote sensing image data has been higher, and the tendency has been seen not only for visible sensor images but also for Synthetic Aperture Radar (SAR) images. Textural analysis has also been carried out for SAR images. However, it is scarcely discussed that textural features of high resolution SAR images such as river, road and residential area are extracted by texture analysis.

Texture analysis refers to acquire texture character through some image processing technology, then obtains quantitative or qualitative description of texture. It includes two aspects: inspecting basic cells of texture and acquiring the information on basic cells arrange distribution of texture. Statistics-based method, structure-based method and spectrum-based method are put forward [15]. Structural texture analysis focuses primarily on identifying periodicity in texture or on identifying their placement rules. Texture analysis has been extensively used to classify remotely sensed images. Filtering features and co-occurrence have been compared in several studies, which concluded that co occurrence features give the best performance. Co-occurrence technique use spatial grey level difference based statistics to extract texture from remote-sensed images. Rignot and Kwok[16] have analyzed SAR images using texture features computed from gray level co occurrence matrices. However, they supplement these features with knowledge about the properties of SAR images. Du[17] used texture features derived from Gabor filters to segment

SAR images. He successfully segmented the SAR images into categories of water, new forming ice, older ice, and multi-year ice. Lee and Philpot (Lee et al,1990) also used spectral texture features to segment SAR images. Various segmentation techniques have been proposed based on statistically measurable features in the image [18-23, 46].Clustering algorithms, such as k-means and ISODATA, operate in an unsupervised mode and have been applied to a wide range of classification problems.

In order to achieve the objective of designing effective algorithms which could provide us with the properties pointed out by Haralick, we think the following ingredients are essential for a textured image segmentation system**:** a set of texture features having good discriminating power; a segmentation algorithm having spatial constraints; and estimation of texture features taking the non stationary nature of the feature image planes into account. We decided to choose a set of existing texture features which can provide us good discriminating power and are easy to compute to serve our need.

The rest of the paper is organized as follows. Section 2 describes proposed algorithm for image segmentation using vector quantization for entropy image. Experimental results for SAR images are in section3 and section 4 concludes the work.

## II. CODEBOOK GENERATION ALGORITHMS

In proposed algorithm we use entropy for grouping pixels into regions and then we form the image of entropy and it is displayed here. To observe the details of an image here we equalized entropy images by histogram equalization method and then applied vector quantization for further segmentation.

### *A. Entropy*

Entropy allows us to consider the neighborhood of the pixel and hence more appropriately texture is considered. Entropy concept was introduced into communications theory by Shannon [24] following the rapid development of communications. It is used to measure the efficiency of the information transferred through a noisy communication channel. The mathematical definition of the entropy given by Shannon is given in equation 1.

$$H = -\sum_{i-1}^{n} P_i \log(P_i) \qquad (1)$$

in which H is the entropy, Pi is the probability of the event i. It defines the average amount of information output of the source generating n symbols. As its magnitude increases, more uncertainty and thus more information is associated with the source. If the source symbols are equally probable, the entropy or uncertainty of Equation 2.1 is maximized and the source provides the greatest possible average information per source symbol.

In this paper probability image is used as an input image to find entropy .Here it becomes necessary to select analyzing window size to find entropy for neighborhood of each pixel in the input image. For this paper 3x3 and 5x5 window sizes were used to find entropy. By moving analyzing window on complete image, calculating entropy for each window, new entropy image was formed by replacing the central pixel of the particular window by entropy and displayed as entropy image. Since these values are very small entropy images were histogram equalized and used for image segmentation as shown in Figure 3(d)

### *B. Vector Quantization (VQ)*

Here to generate two vectors v1 & v2 proportionate error is Vector Quantization (VQ) [25-31, 47] is an efficient technique for data compression and has been successfully used in various applications such as speech data compression [32], content based image retrieval CBIR [33]. VQ has been very popular in a variety of research fields such as speaker recognition and face detection [34, 35]. VQ is also used in real time applications such as real time video-based event detection [36] and anomaly intrusion detection systems [37], image segmentation [38-41, 48-51] and face recognition [42].

VQ is a technique in which a codebook is generated for each image. A codebook is a representation of the entire image containing a definite pixel pattern which is computed according to a specific VQ algorithm. The image is divided into fixed sized blocks that form the training vector. The generation of the training vector is the first step to cluster formation. On these training vectors clustering methods are applied and codebook is generated. The method most commonly used to generate codebook is the Linde Buzo Gray (LBG) algorithm [43].

### *C. Kekre's Fast Codebook Genration Algorithm (KFCG)*

Here the Kekre's Fast Codebook Generation algorithm [44,45] proposed for image data compression is used. This algorithm reduces the time for code book generation. Initially we have one cluster with the entire training vectors and the codevector C1 which is centroid. In the first iteration of the algorithm, the clusters are formed by comparing first element of training vector with first element of code vector C1. The vector Xi is grouped into the cluster 1 if xi1< c11 otherwise vector Xi is grouped into cluster2 as shown in Figure 1(a) where codevector dimension space is 2. In second iteration, the cluster 1 is split into two by comparing second element xi2 of vector Xi belonging to cluster 1 with that of the second element of the codevector. Cluster 2 is split into two by comparing the second element xi2 of vector Xi belonging to cluster 2 with that of the second element of the codevector as shown in Figure 1(b).

This procedure is repeated till the codebook size is reached to the size specified by user. It is observed that this algorithm gives less error as compared to LBG and requires least time to generate codebook as compared to other algorithms, as it does not require any computation of Euclidean distance. The algorithm shown in Figure 1(a) and Figure 1(b) for two dimensional case it is easily extended to higher dimensions.

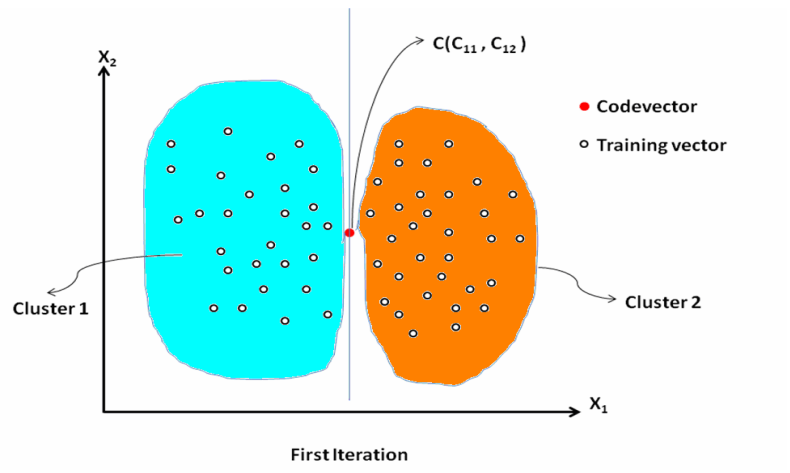

Figure 1(a)

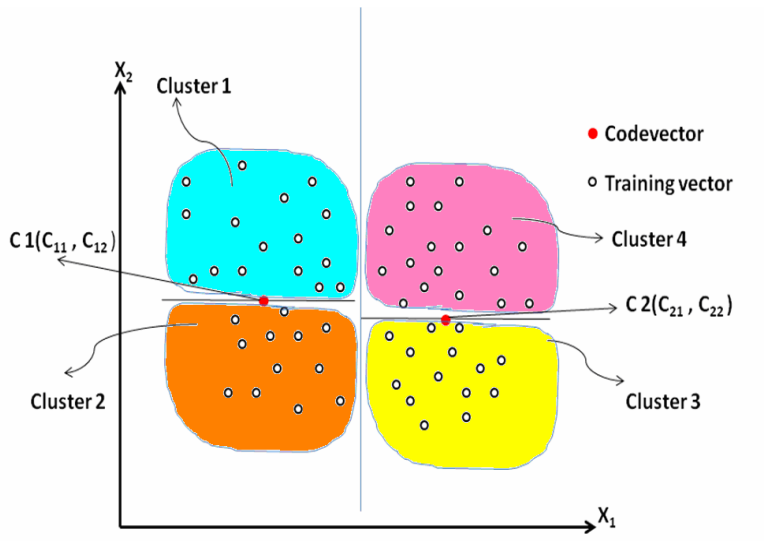

Figure 1b

Figure 1 : KFCG algorithm for 2 dimensional case

In this paper initially codebook of size 128 using 4 dimensional vector space was selected. Thus the image is divided into 128 clusters which were further reduced to 8 by using requantization. The 8 clusters thus obtained were mapped onto the image generating 8 different images representing them. On all these images Canny's operator was used to obtain the edge maps. These edge maps were superimposed on the original image giving proper segmentation of image as shown in Figure 4(a) and (b).

## III. Results

Here image of Vihar lake (Figure 2(a)) is used to test proposed algorithm. First entropy was calculated from probability for original image. As proceed further with histogram equalization for probability and entropy, we can differentiate between different texture, displayed in Figure 2(b) & (d). Images were tested for analyzing window size of 3x3 and 5x5 to find entropy. Results for window size3x3 are displayed here. In third step KFCG algorithm was used on equalized entropy image for further segmentation.

Using equalized entropy image as an input image codebook of size 128 using KFCG algorithm was generated. Further these code-vectors were clustered in 8 clusters using same algorithm. Then 8 images were constructed using one code-vector at a time as shown in Figure 3(a)-(h) for KFCG algorithm. For comparison we displayed results of Entropy and equalized entropy using GLCM as shown in Figure 5(a) and (b). Figure 4(a) indicates segmented roads and Figure 4(b) shows segmented water where edge map is superimposed on original image for codevector four and eight reclustered images.

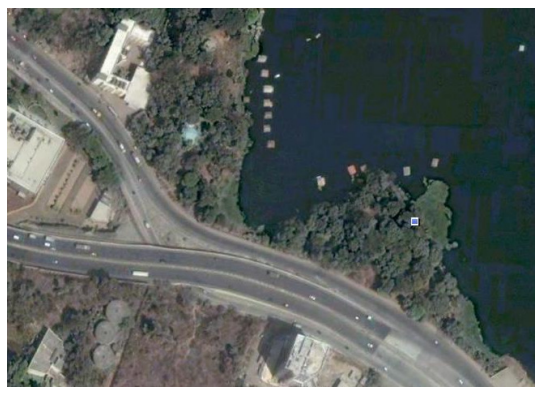

(a)Original Image

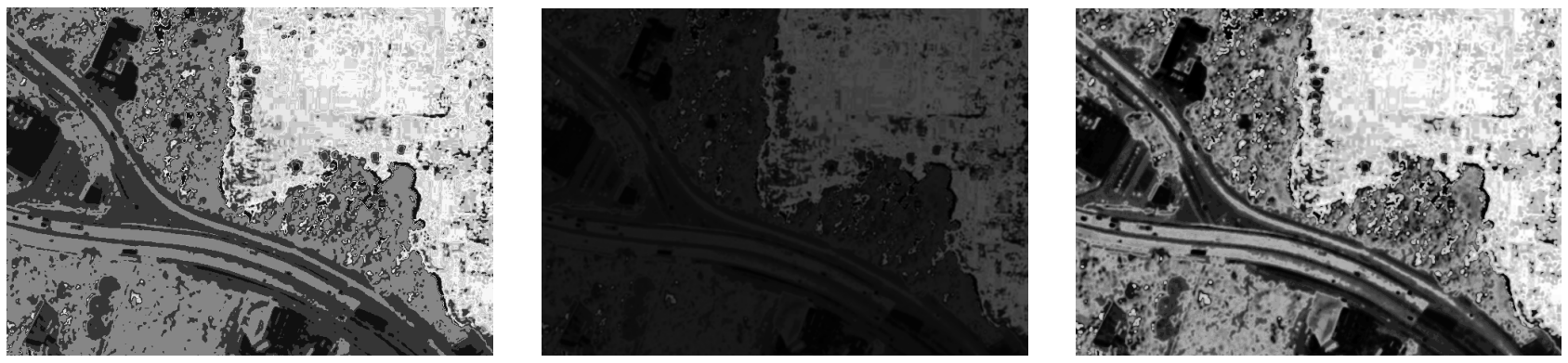

(b) (c) (d)

Figure 2: (a) Original image, (b)Histogram equalized probability image, (c) Direct Entropy image for window size (3x3),(d)Histogram equalized image for (c)

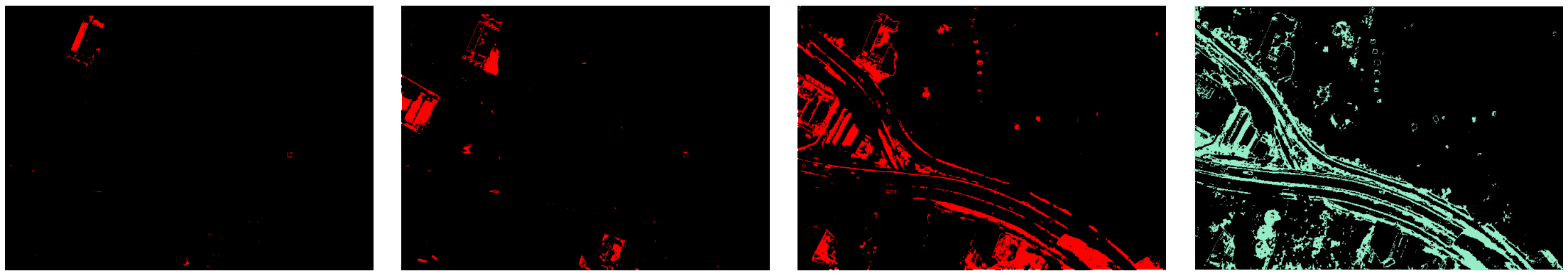

(a) (b) (c) (d)

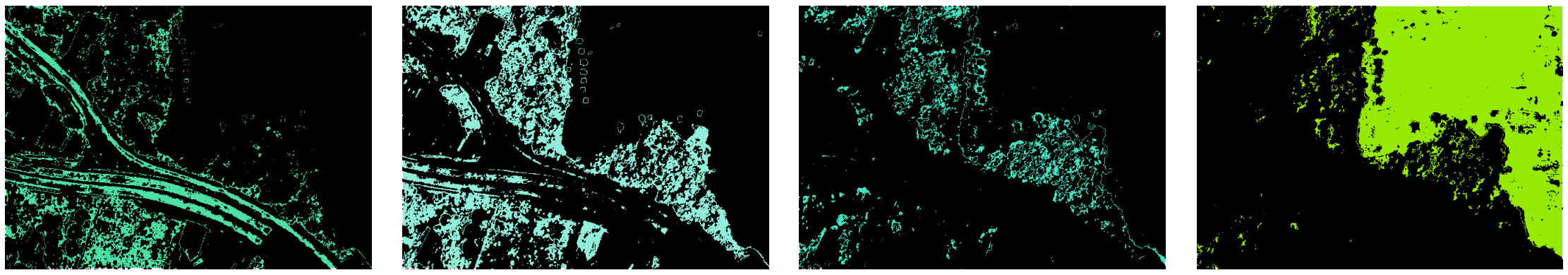

(e) (f) (g) (h)

Figure 3: (a) Image for first code-vector ,(b) Image for second code-vector, (c)Image for third code-vector, (d)Image for fourth code-vector, (e)Image for fifth code-vector, (f)Image for sixth code-vector, (g)Image for seventh code-vector, (h)Image for eighth code-vector

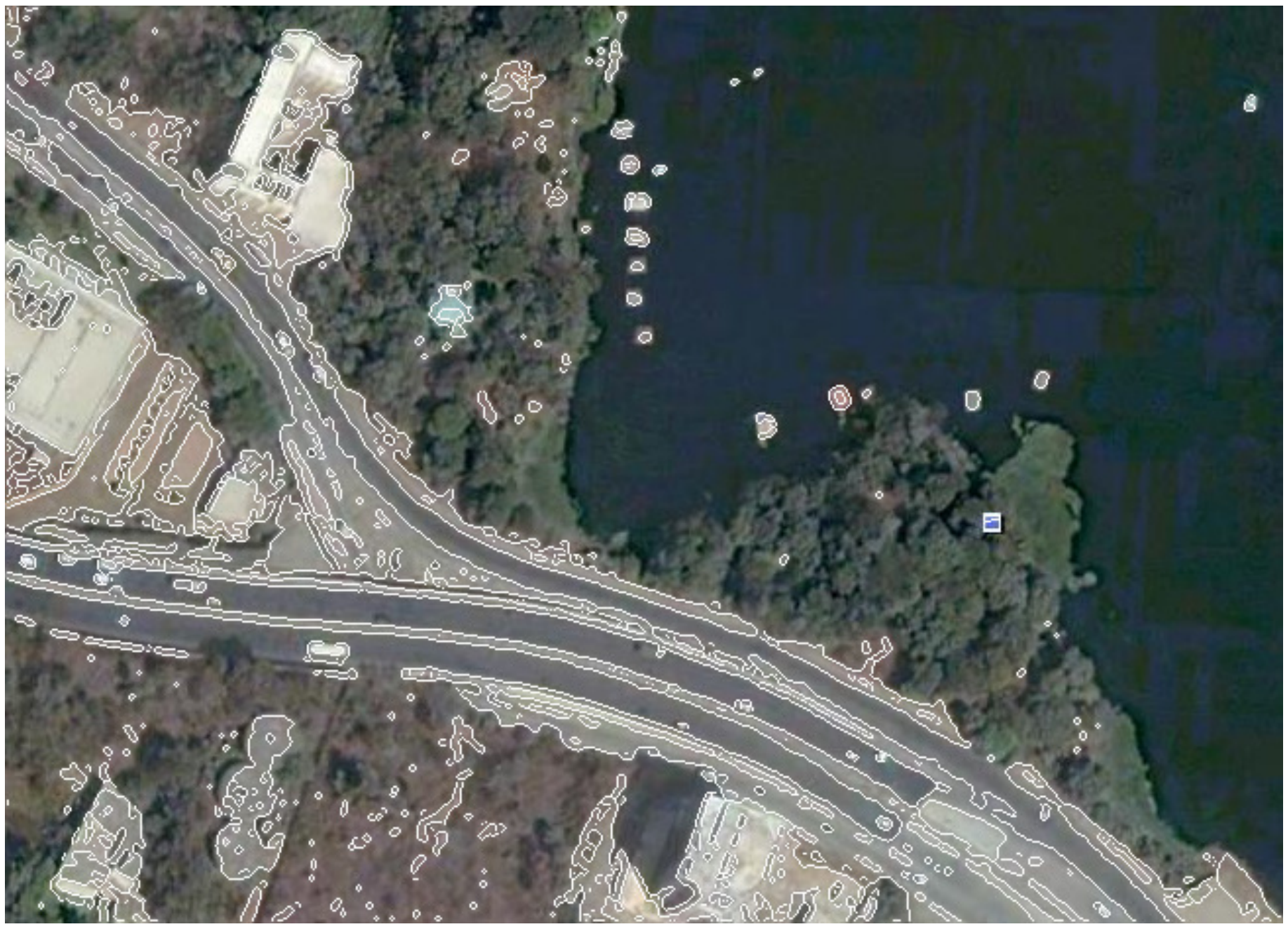

(a)

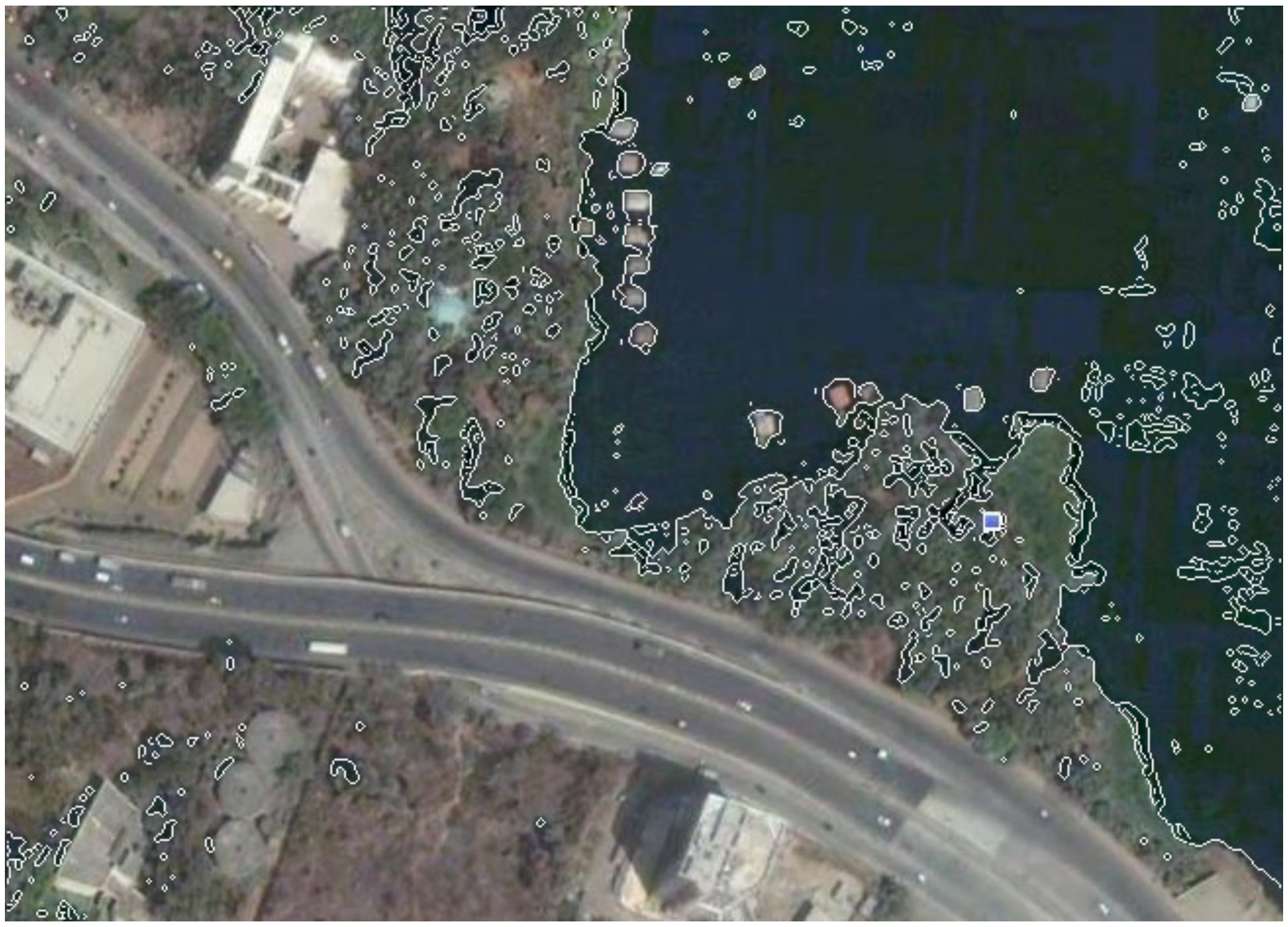

(b)

Figure 4: (a) Superimposed image for forth codevector, (b) Superimposed image for eighth codevector.

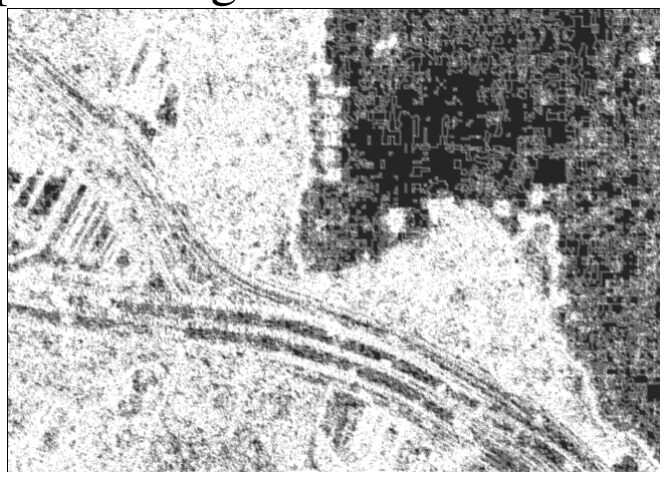

(a)

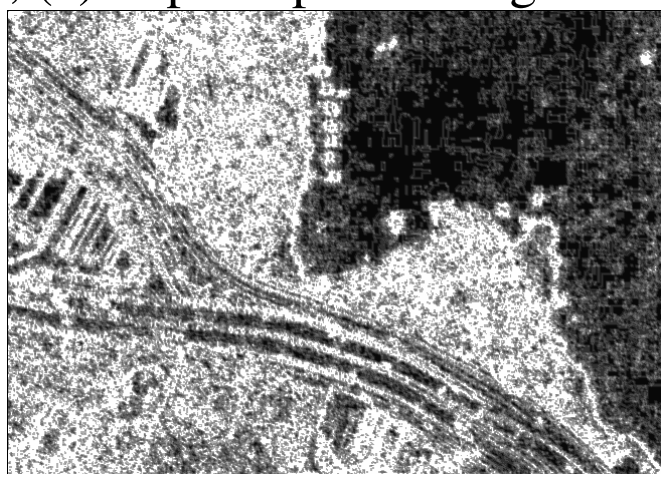

(b)

Figure 5: (a)Entropy for Figure2(a) using GLCM ,(b) Equalized Entropy for Figure 5(a)

## V. CONCLUSION

In this paper a new segmentation method for SAR image is proposed. The proposed segmentation uses clustering algorithm Kekre's Fast Codebook Generation (KFCG) which is a vector quantization technique. Initially, entropy image is obtained and then clustering of entropy image is done using KFCG algorithm for segmentation purpose. The codebook of size 128 is generated for the Entropy image. These code vectors were further clustered to 8 clusters using same KFCG algorithm. The 8 images corresponding to each codevector is displayed in the result section. It is observed that in Figure 3(b) corresponding to second codevector buildings gets segmented, image corresponding to forth codevector i.e. Figure 3(d) segments buildings and roads where as Figure 3(h) corresponding to the last codevector segments the water area from the original image. After comparing the proposed algorithm with GLCM results given in Figure 5(a) and (b) ,proposed method gives far better segmentation.

## AUTHOR BIOGRAPHIES

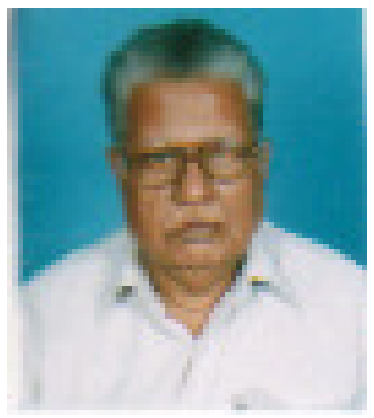

**Dr. H. B. Kekre** has received B.E. (Hons.) in Telecomm. Engg. from Jabalpur University in 1958, M.Tech (Industrial Electronics) from IIT Bombay in 1960, M.S.Engg. (Electrical Engg.) from University of Ottawa in 1965 and Ph.D. (System Identification) from IIT Bombay in 1970. He has worked Over 35 years as Faculty of Electrical Engineering and then HOD Computer Science and Engg. at IIT Bombay. For last 13 years worked as a Professor in Department of Computer Engg. at Thadomal Shahani Engineering College, Mumbai. He is currently Senior Professor working with Mukesh Patel School of Technology Management and Engineering, SVKM’s NMIMS University, Vile Parle(w), Mumbai, INDIA. His areas of interest are Digital Signal processing and Image Processing. He has more than 250 papers in National / International Conferences / Journals to his credit. Recently six students working under his guidance have received best paper awards. Currently he is guiding ten Ph.D. students.

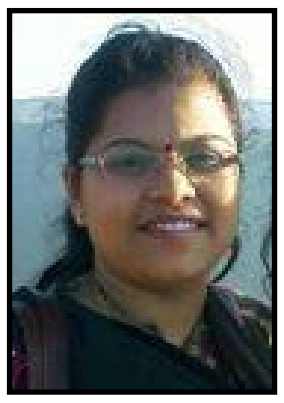

**Ms. Saylee M. Gharge** has received M.E. (Electronics and telecomm.) degree from Mumbai University in 2007, currently Pursuing Ph.D. from Mukesh Patel School of Technology, Management and Engineering, SVKM’s NMIMS University, Vile-Parle (W), Mumbai. She has more than 10 years of experience in teaching. Currently working as a Assistant Professor in department of electronics and telecommunication in Vivekanand Institute of Technology, Mumbai. Her areas of interest are Image Processing, Signal Processing. She has 30 papers in National /International Conferences/journal to her credit.

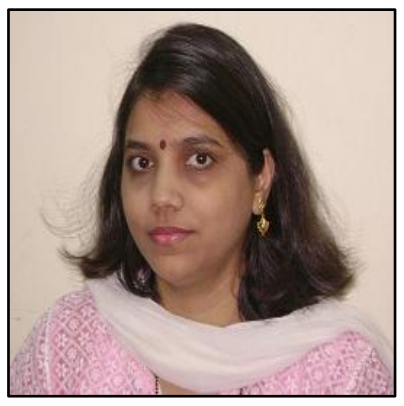

**Tanuja K. Sarode** has Received Bsc.(Mathematics) from Mumbai University in 1996, Bsc.Tech.(Computer Technology) from Mumbai University in 1999, M.E. (Computer Engineering) degree from Mumbai University in 2004, currently Pursuing Ph.D. from Mukesh Patel School of Technology, Management and Engineering, SVKM’s NMIMS University, Vile-Parle (W), Mumbai, INDIA. She has more than 10 years of experience in teaching. Currently working as Associate Professor in Dept. of Computer Engineering at Thadomal Shahani Engineering College, Mumbai. She is life member of IETE, member of International Association of Engineers (IAENG) and International Association of Computer Science and Information Technology (IACSIT), Singapore. Her areas of interest are Image Processing, Signal Processing and Computer Graphics. She has 55 papers in National /International Conferences/journal to her credit.